\documentclass[twocolumn]{aastex63}

\received{2020 April 7}
\revised{2020 April 22}
\accepted{2020 April 25}
\published{2020 June 8}
\submitjournal{ApJL}

\usepackage{multirow}
\usepackage{txfonts}
\usepackage{graphicx}
\usepackage[outdir=./]{epstopdf}
\usepackage{natbib}
\usepackage{xcolor, color}
\usepackage{booktabs}
\usepackage{threeparttable}
\usepackage{hyperref}
\hypersetup{linkcolor=red,citecolor=blue,filecolor=cyan,urlcolor=magenta}

\shortauthors{De Simone et al.}

\begin{document}

\title{Hot Corinos chemical diversity: myth or reality? }

\correspondingauthor{Marta De Simone}
\email{marta.desimone@univ-grenoble-alpes.fr}

\author[0000-0001-5659-0140]{Marta De Simone}
\affiliation{Univ. Grenoble Alpes, CNRS, IPAG, 38000 Grenoble, France}

\author[0000-0001-9664-6292]{Cecilia Ceccarelli}
\affiliation{Univ. Grenoble Alpes, CNRS, IPAG, 38000 Grenoble, France}

\author[0000-0003-1514-3074]{Claudio Codella}
\affiliation{INAF, Osservatorio Astrofisico di Arcetri, Largo E. Fermi 5, 50125 Firenze, Italy}
\affiliation{Univ. Grenoble Alpes, CNRS, IPAG, 38000 Grenoble, France}

\author[0000-0002-8502-6431]{Brian E. Svoboda}
\affiliation{National Radio Astronomy Observatory\footnote{\label{fn:NRAO}The National Radio Astronomy Observatory is a facility of the National Science Foundation operated under cooperative agreement by Associated Universities, Inc.}, 1003 Lopezville Rd, Socorro, NM 87801, USA}
\affiliation{Steward Observatory, University of Arizona, 933 North Cherry Avenue, Tucson, AZ 85721, USA}

\author[0000-0002-7570-5596]{Claire Chandler}
\affiliation{National Radio Astronomy Observatory\footnote{\label{fn:NRAO}The National Radio Astronomy Observatory is a facility of the National Science Foundation operated under cooperative agreement by Associated Universities, Inc.}, 1003 Lopezville Rd, Socorro, NM 87801, USA}

\author[0000-0003-0167-0746]{Mathilde Bouvier}
\affiliation{Univ. Grenoble Alpes, CNRS, IPAG, 38000 Grenoble, France}

\author[0000-0002-9865-0970]{Yamamoto Satoshi}
\affiliation{Department of Physics, The University of Tokyo, Bunkyo-ku, Tokyo 113-0033, Japan}

\author[0000-0002-3297-4497]{Nami Sakai}
\affiliation{The Institute of Physical and Chemical Research (RIKEN), 2-1, Hirosawa, Wako-shi, Saitama 351-0198, Japan}

\author[0000-0003-1481-7911]{Paola Caselli}
\affiliation{Max-Planck-Institut f\"ur extraterrestrische Physik (MPE), Giessenbachstrasse 1, 85748 Garching, Germany}

\author[0000-0002-5789-6931]{Cecile Favre}
\affiliation{Univ. Grenoble Alpes, CNRS, IPAG, 38000 Grenoble, France}

\author[0000-0002-5635-3345]{Laurent Loinard}
\affiliation{Instituto de Radioastronomía y Astrofísica, Universidad Nacional Autónoma de México Apartado 58090, Morelia, Michoacán, Mexico}

\author[0000-0002-9397-3826]{Bertrand Lefloch}
\affiliation{Univ. Grenoble Alpes, CNRS, IPAG, 38000 Grenoble, France}

\author[0000-0003-2300-2626]{Hauyu Baobab Liu}
\affiliation{Academia Sinica Institute of Astronomy and Astrophysics (ASIAA), No. 1, Section 4, Roosevelt Road, Taipei 10617, Taiwan}

\author[0000-0002-6729-3640]{Ana L\'opez-Sepulcre}
\affiliation{Institut de Radioastronomie Millim\'etrique (IRAM), 300 rue de la Piscine, 38400 Saint-Martin d'H\`eres, France}
\affiliation{Univ. Grenoble Alpes, CNRS, IPAG, 38000 Grenoble, France}

\author[0000-0002-3972-1978]{Jaime E. Pineda}
\affiliation{Max-Planck-Institut f\"ur extraterrestrische Physik (MPE), Giessenbachstrasse 1, 85748 Garching, Germany}

\author[0000-0002-2308-2585]{Vianney Taquet}
\affiliation{INAF, Osservatorio Astrofisico di Arcetri, Largo E. Fermi 5, 50125 Firenze, Italy}

\author[0000-0003-1859-3070]{Leonardo Testi}
\affiliation{ESO, Karl Schwarzchild Srt. 2, 85478 Garching bei M\"unchen, Germany
}
\affiliation{INAF, Osservatorio Astrofisico di Arcetri, Largo E. Fermi 5, 50125 Firenze, Italy}
\affiliation{Excellence Cluster Origins, Boltzmannstrasse 2, D-85748 Garching bei München, Germany}


\begin{abstract}
After almost 20 years of hunting, only about a dozen hot corinos, hot regions enriched in interstellar complex organic molecules (iCOMs), are known. Of them, many are binary systems with the two components showing drastically different molecular spectra. Two obvious questions arise. Why are hot corinos so difficult to find and why do their binary components seem chemically different? The answer to both questions could be a high dust opacity that would hide the molecular lines. To test this hypothesis, we observed methanol lines at centimeter wavelengths, where dust opacity is negligible, using the Very Large Array interferometer. We targeted the NGC 1333 IRAS 4A binary system, for which one of the two components, 4A1, has a spectrum deprived of iCOMs lines when observed at millimeter wavelengths, while the other component, 4A2, is very rich in iCOMs. We found that centimeter methanol lines are similarly bright toward 4A1 and 4A2. 
Their non-LTE analysis indicates gas density and temperature ($\geq2\times10^6$ cm$^{-3}$ and 100--190 K), methanol column density ($\sim10^{19}$ cm$^{-2}$) and extent ($\sim$35 au in radius) similar in 4A1 and 4A2, proving that both are hot corinos. Furthermore, the comparison with previous methanol line millimeter observations allows us to estimate the optical depth of the dust in front of 4A1 and 4A2, respectively. The obtained values explain the absence of iCOMs line emission toward 4A1 at millimeter wavelengths and indicate that the abundances toward 4A2 are underestimated by $\sim$30\%.
Therefore, centimeter observations are crucial for the correct study of hot corinos, their census, and their molecular abundances.
\end{abstract}


\section{Introduction} \label{sec:intro}
Interstellar complex organic molecules (iCOMs) are molecules detected in the interstellar medium containing carbon and
at least six atoms \citep{herbst_complex_2009, ceccarelli_seeds_2017}. These molecules are of particular interest because
they carry a substantial fraction of carbon that can be used for prebiotic chemistry \citep[e.g.][]{caselli_our_2012}. 

In solar-like young Class 0 protostars, iCOMs are found in relatively large quantities toward the so-called hot corinos, which are compact ($\leq$100 au), hot ($\geq$100 K) and dense ($\geq10^7$ cm$^{-3}$) regions enriched in iCOMs at the center of the envelopes accreting the future star \citep{ceccarelli_hot_2004,ceccarelli_extreme_2007,caselli_our_2012}. 

The first hot corino was discovered in 2003 toward the Class 0 source IRAS 16293--2422 \citep[e.g.][]{cazaux_hot_2003,jorgensen_alma_2016, manigand_alma-pils_2020}. Since then other Class 0 hot corinos have been discovered: NGC 1333 IRAS 4A \citep[hereafter IRAS 4A; e.g.][]{bottinelli_complex_2004,taquet_constraining_2015,lopez-sepulcre_complex_2017,de_simone_glycolaldehyde_2017,sahu_implications_2019}, NGC 1333 IRAS 2A, NGC 1333 IRAS 4B \citep[e.g. ][]{jorgensen_probing_2005,bottinelli_hot_2007,maury_first_2014,de_simone_glycolaldehyde_2017}, HH 212 \citep{codella_water_2016,bianchi_deuterated_2017,lee_formation_2017,lee_first_2019}, B335 \citep{imai_discovery_2016}, L483 \citep{oya_l483_2017,jacobsen_organic_2018}, Barnard1b-S \citep{marcelino_alma_2018}, Ser-emb 1 \citep{martin-domenech_new_2019}, BHR71-IRS1 \citep{yang_constraining_2020}. 
Lately, few more evolved Class I hot corinos were also discovered: NGC 1333 SVS13A \citep{de_simone_glycolaldehyde_2017,bianchi_census_2019}, B1a \citep{oberg_complex_2014} and Ser-emb 17 \citep{bergner_organic_2019}.
Therefore, after almost 20 years, only about a dozen hot corinos are known. Recent surveys concluded that $\sim$30\% of low-mass Class 0/I protostars show emission from at least three iCOMs \citep{de_simone_glycolaldehyde_2017,belloche_questioning_2020}. 

Most of the hot corinos cited above turn out to be binary systems when imaged at high angular resolution. This is in agreement with previous surveys that found that 40-60\% of protostars are multiple systems \citep{maury_first_2014,tobin_vla_2016.} 
Interestingly, with the first hot corino maps it became clear that the two objects in a given binary system can substantially differ in molecular complexity. 
Illustrative examples are provided by IRAS 16293--2422 and IRAS 4A \citep{jorgensen_alma_2016, lopez-sepulcre_complex_2017}. 
IRAS 16293--2422 is composed by two sources, A and B, separated by $5\farcs1$ ($\sim$ 720 au), where source A, weaker in millimeter continuum emission, is brighter in iCOMs lines than source B \citep[e.g.][]{caux_timasss_2011,pineda_first_2012,jorgensen_alma_2016,manigand_alma-pils_2020}.
IRAS 4A, located in the NGC 1333 region in the Perseus cloud at $(299\pm15)$ pc of distance \citep{zucker_mapping_2018}, is also a binary system composed by IRAS 4A1 and IRAS 4A2 (hereafter 4A1 and 4A2), separated by 1.8$''$ ($\sim$540 au): while 4A1 is brighter in the mm continuum than 4A2, only 4A2 shows bright iCOMs lines \citep{taquet_constraining_2015,lopez-sepulcre_complex_2017,de_simone_glycolaldehyde_2017}. 
However, not always the brightest millimeter continuum source in a binary system is the one weak in iCOMs emission \citep[see e.g.][]{ospina-zamudio_first_2018}.

In summary, despite two decades of hunting, only a dozen hot corinos are known so far. 
Of them, many are binary systems with the two components showing drastically different molecular spectra. 
Two related questions arise: (1) Why are hot corinos so difficult to find? While it is known that not all Class 0/I sources possess hot corinos \citep[e.g.][]{sakai_warm_2013, higuchi_chemical_2018, bouvier_hunting_2020}, observational biases might hamper their detection. 
(2) Why do coeval objects seem drastically differ in their chemical composition? Is this a real difference or is it only/mostly due to observational biases?

A major observational bias could be caused by the dust opacity, which could be very high in Class 0/I sources, due to their high densities and, consequently, column densities \citep[e.g.][]{miotello_grain_2014,galvan-madrid_effects_2018,galametz_low_2019}. 
If the effect of dust absorption is not negligible, there are three major consequences: (1) hot corinos may be difficult to detect in the millimeter (also) because of the high dust absorption of the iCOMs lines; (2) the molecular complexity diversity observed in binary systems objects may reflect a difference in the front dust column density rather than a real chemical difference of the two objects; (3) the iCOMs abundances in hot corinos could have been so far underestimated.
In order to test this hypothesis, we targeted the IRAS 4A binary system, where the two objects show extremely different iCOMs line spectra at mm wavelengths (see above), and carried out observations of several methanol lines, one of the simplest iCOMS, at centimeter wavelengths where the dust is optically thin.

\section{Observations} \label{sec:obs}
The IRAS 4A system was observed at 1.3 cm using K--band receivers (18-26.5 GHz) of the Very Large Array (VLA) in C-configuration (35--3400 m) on 2018 December 10 (project ID: VLA/18B-166). 
We targeted 10 CH$_3$OH lines, with frequencies from 24.9 to 26.4 GHz, upper level energies $E_{up}$ from 36 to 175 K and Einstein coefficients $A_{ij}$ in the range (0.5 -- 1.1) $\times 10^{-7}$ s$^{-1}$ (Table \ref{tab:spectral_params&fit_res}). 
The observed spectra were divided into eight spectral windows with $\sim$0.017 MHz (0.2 km s$^{-1}$) spectral resolution and $\sim1''$ ($\sim$300 au at the distance of IRAS 4A) angular resolution. 
The observations were centered on 4A2, at $\rm \alpha(J2000)=03^h29^m10\fs43$, $\rm \delta(J2000)=31^\circ13'32\farcs1$. The flux calibrators were J0137+3309 and J0521+1638, while the bandpass and the gain ones were J0319+4130, and J0336+3218, respectively. The absolute flux calibration error is $\leq$15\%\footnote{https://science.nrao.edu/facilities/vla/docs/manuals/oss/performance/fdscale}.

The data reduction and cleaning process were performed using the CASA\footnote{https://casa.nrao.edu/} package while data analysis and images were performed using the GILDAS\footnote{http://www.iram.fr/IRAMFR/GILDAS} package. 
We obtained a continuum image by averaging line-free channels from all the spectral windows (Figure \ref{fig:methanol_maps+cont}). 
We self-calibrated, in phase amplitude, using the line-free continuum channels and applied the solutions to both the continuum and molecular lines. The dynamic range, as defined by peak source flux over rms noise, was improved by 20\% by the self-calibration. The final RMS noise in the continuum image, 3$\mu$Jy beam$^{-1}$, is consistent with that reported by the VLA Exposure Time Calculator for a line-free continuum bandwidth of 4.5GHz, 26 antennas, and an on-source integration time of 3 hr.
The cube were subsequently continuum subtracted, smoothed to 1 km s$^{-1}$ ($\sim$ 0.08 MHz) and cleaned in CASA using a multiscale deconvolution\footnote{
This technique is a scale-sensitive deconvolution algorithm efficient for images with complicated and extended spatial structures. In fact, it allows us to model the sky brightness as a linear combination of flux components of different scale sizes. The scale sizes are chosen following approximately the sizes of the dominant structures in the image and including the ``0'' scale to
model the unresolved ones (see \href{https://casa.nrao.edu/casadocs/casa-5.1.0/synthesis-imaging/deconvolution-algorithms}{casadocs-deconvolution-algorithms}).
} (scales=[0,5,15,18,25]) with natural weighting. 
The synthesized beams for each spectral window are reported in Table \ref{tab:spectral_params&fit_res}. The half power primary beam is $\sim 80''$.

\begin{table*}
    \centering
    \resizebox{\textwidth}{!}{%
    \begin{threeparttable}
        \caption{Spectral parameters, synthesized beams and Gaussian fit results of the CH$_3$OH lines extracted toward the 4A1 and 4A2 continuum peaks.
        }\label{tab:spectral_params&fit_res}
        \begin{tabular}{l|cccc|cccc|cccc}
            \hline
            \hline
            \multirow{2}{*}{Transition} & \multirow{2}{*}{Frequency$^{(a)}$} & \multirow{2}{*}{E$_{\rm up}^{(a)}$}& \multirow{2}{*}{logA$_{\rm ij}^{(a)}$} & Synthesized Beam & \multicolumn{4}{c|}{IRAS 4A1} & \multicolumn{4}{c}{IRAS 4A2}    \\
                &   &   &   & maj $\times$ min (PA)  &   $\rm \int T_BdV^b$ & V$_{\rm peak}^b$ & FWHM$^b$ & RMS$^c$ & $\rm \int T_BdV^b$ & V$_{\rm peak}^b$ & FWHM$^b$ & RMS$^c$ \\ 
                & [GHz] & [K] &  & [$''\times ''$ ($^\circ$)] & [K km s$^{-1}$] & [km s$^{-1}$] & [km s$^{-1}$] & [K] & [K km s$^{-1}$] & [km s$^{-1}$] & [km s$^{-1}$] & [K] \\
            \hline
            3(2,1)-3(1,2) E & 24.92871 & 36 & -7.2 & 
            $0.97\times0.95 \ (-12)$ & 17(4) & 6.5(0.2) & 2.7(1.2) & 1.0 & 34(3) & 6.8(0.2) & 3.1(0.3) & 0.9 \\
            4(2,2)-4(1,3) E & 24.93347 & 45 & -7.1 & $0.97\times0.95 \ (-12)$ & 23(3) & 6.5(0.2) & 3.7(0.5) & 1.0 & 32(3) & 6.9(0.2) & 3.1(0.3) & 0.9 \\
            2(2,0)-2(1,1) E & 24.93438 & 29 & -7.2 & $0.97\times0.95 \ (-12)$ & 19(3) & 6.5(0.2) & 3.3(0.6) & 1.0 & 27(3) & 6.9(0.2) & 2.9(0.3) & 0.9 \\
            5(2,3)-5(1,4) E & 24.95908 & 57 & -7.1 & $0.97\times0.95 \ (-12)$ & 19(3) & 6.1(0.3) & 4.2(0.9) & 0.9 & 32(3) & 6.8(0.2) & 3.1(0.3) & 0.9 \\
            6(2,4)-6(1,5) E & 25.01812 & 71 & -7.1 & $0.97\times0.95 \ (-19)$ & 19(3) & 6.3(0.3) & 3.5(0.6) & 1.0 & 31(3) & 6.7(0.1) & 2.5(0.2) & 1.0 \\
            7(2,5)-7(1,6) E & 25.12487 & 87 & -7.1 & $0.98\times0.95 \ (-21)$ & 20(2) & 7.0(0.2) & 3.2(0.5) & 0.9 & 35(3) & 6.8(0.2) & 2.9(0.3) & 1.2 \\
            8(2,6)-8(1,7) E & 25.29442 & 106 & -7.0 & $0.96\times0.94 \ (-11)$ & 13(3) & 6.8(0.3) & 3.4(0.9) & 0.9 & 32(2) & 6.8(0.1) & 2.8(0.3) & 0.9 \\
            9(2,7)-9(1,8) E & 25.54140 & 127 & -7.0 & $0.96\times0.92 \ (-50)$ & 18(2) & 6.5(0.2) & 2.7(0.5) & 0.8 & 31(2) & 6.9(0.1) & 2.6(0.2) & 0.9 \\
            10(2,8)-10(1,9) E & 25.87827 & 150 & -7.0 & $0.97\times0.93 \ (-35)$ & 20(2) & 6.4(0.2) & 3.2(0.5) & 0.8 & 32(2) & 6.9(0.1) & 2.6(0.2) & 0.8 \\
            11(2,9)-11(1,10) E & 26.31312 & 175 & -6.9 & $0.94\times0.91 \ (-35)$ & 24(4) & 6.0(0.4) & 4.9(0.9) & 1.1 & 31(3) & 6.8(0.1) & 2.9(0.3) & 0.9 \\
            \hline
            \end{tabular}
        \begin{tablenotes}
            \item[a] Spectroscopic parameters by \citet{xu_torsion_2008} are from the CDMS \citep[Cologne Database for Molecular Spectroscopy;][]{muller_cologne_2005} molecular database.
            \item[b] Results of the Gaussian fit algorithm.
            \item[c] The RMS is computed over each spectral window.
        \end{tablenotes}
    \end{threeparttable}
    }
\end{table*}

\section{Results} \label{sec:results}

\subsection{Continuum emission map}
Figure \ref{fig:methanol_maps+cont} reports the map of the continuum emission at 25 GHz. The two continuum peaks mark the two protostars, whose coordinates ($\rm \alpha(J2000)=03^h29^m10\fs536$, $\rm \delta(J2000)=31^\circ13'31\farcs07$ for 4A1, and $\rm \alpha(J2000)=03^h29^m10\fs43$, $\rm \delta(J2000)=31^\circ13'32\farcs1$ for 4A2) are consistent with those derived by \citet{tobin_vla_2016} and \citet{lopez-sepulcre_complex_2017} with higher angular resolution observations. 
Since the angular resolution of our observations ($\sim1''$) is smaller than the separation between 4A1 and 4A2 (1$'\farcs$8), they are clearly disentangled in our images, even if individually unresolved with the current resolution.

At cm wavelengths, 4A1 shows a brighter continuum emission (due to dust or free-free) than 4A2.  
The peak fluxes are (2.1 $\pm$ 0.3) mJy beam$^{-1}$ and (0.47 $\pm$ 0.07) mJy beam$^{-1}$ toward 4A1 and 4A2, respectively. Taking into account the slightly different wavelength (1.05 cm) and angular resolution ($\mathbf{\sim 0\farcs1}$), these values are consistent with the ones measured by \citet{tobin_vla_2016}: (1.3 $\pm$ 0.2) mJy beam$^{-1}$ for 4A1 and (0.38 $\pm$ 0.04) mJy beam$^{-1}$ for 4A2.

\subsection{Methanol lines}
\begin{figure*}
    \centering
    \includegraphics[scale=0.5]{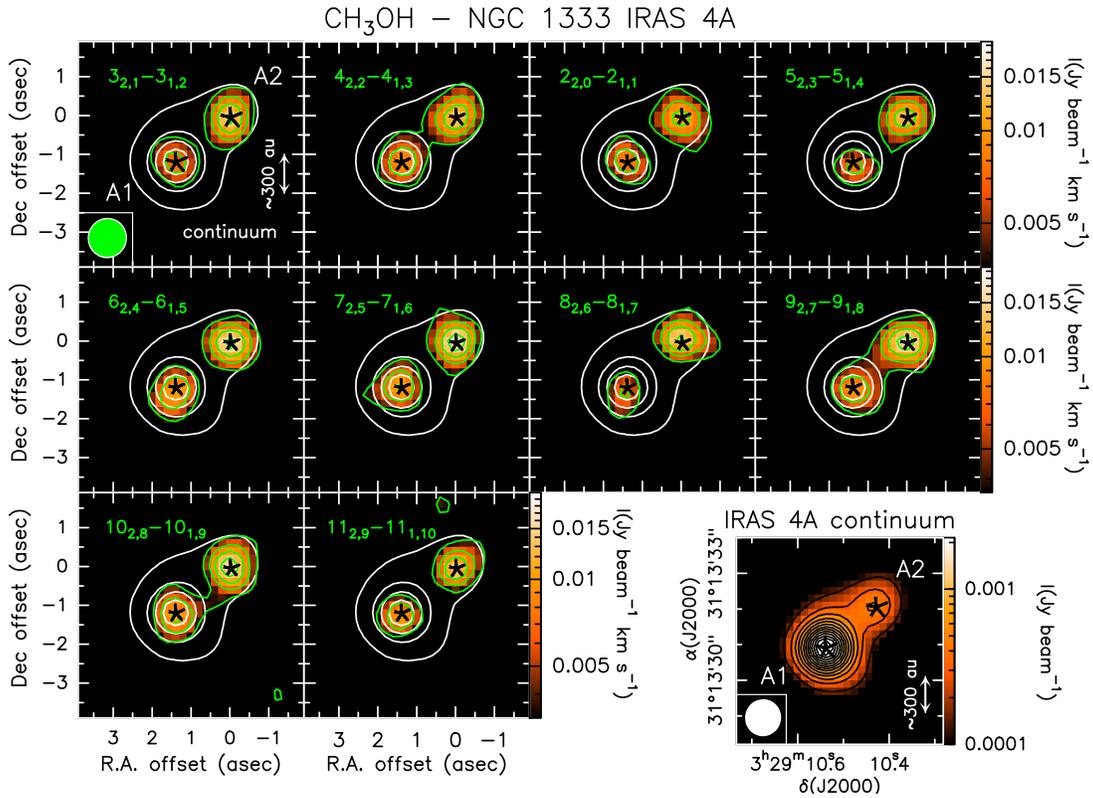}
    \caption{\textit{Bottom right panel:} IRAS 4A continuum emission map at 25 GHz. First contour and steps correspond to 50$\rm \sigma_C$ ($\rm \sigma_C=3\mu$Jy beam$^{-1}$).
    \textit{Other panels:} CH$_3$OH velocity-integrated maps toward IRAS 4A in color scale overlapped with the continuum (white) contours (from 50$\rm \sigma_C$ with steps of 170$\rm \sigma_C$). The emission is integrated from -2 km s$^{-1}$ to 2 km s$^{-1}$ with respect to the $\rm v_{sys}$ ($\sim6.7$ km s$^{-1}$). Methanol first contour (green) and steps correspond to 3$\sigma$ ($\sigma$=1.2 mJy beam$^{-1}$ km s$^{-1}$). The transition of the imaged line is reported in each panel.
    The black stars show the 4A1 and 4A2 positions. Synthesised beams for continuum (white) and lines (green) are in the lower left corner.}
    \label{fig:methanol_maps+cont}
\end{figure*}
\begin{figure*}
    \centering
    \includegraphics[scale=0.7]{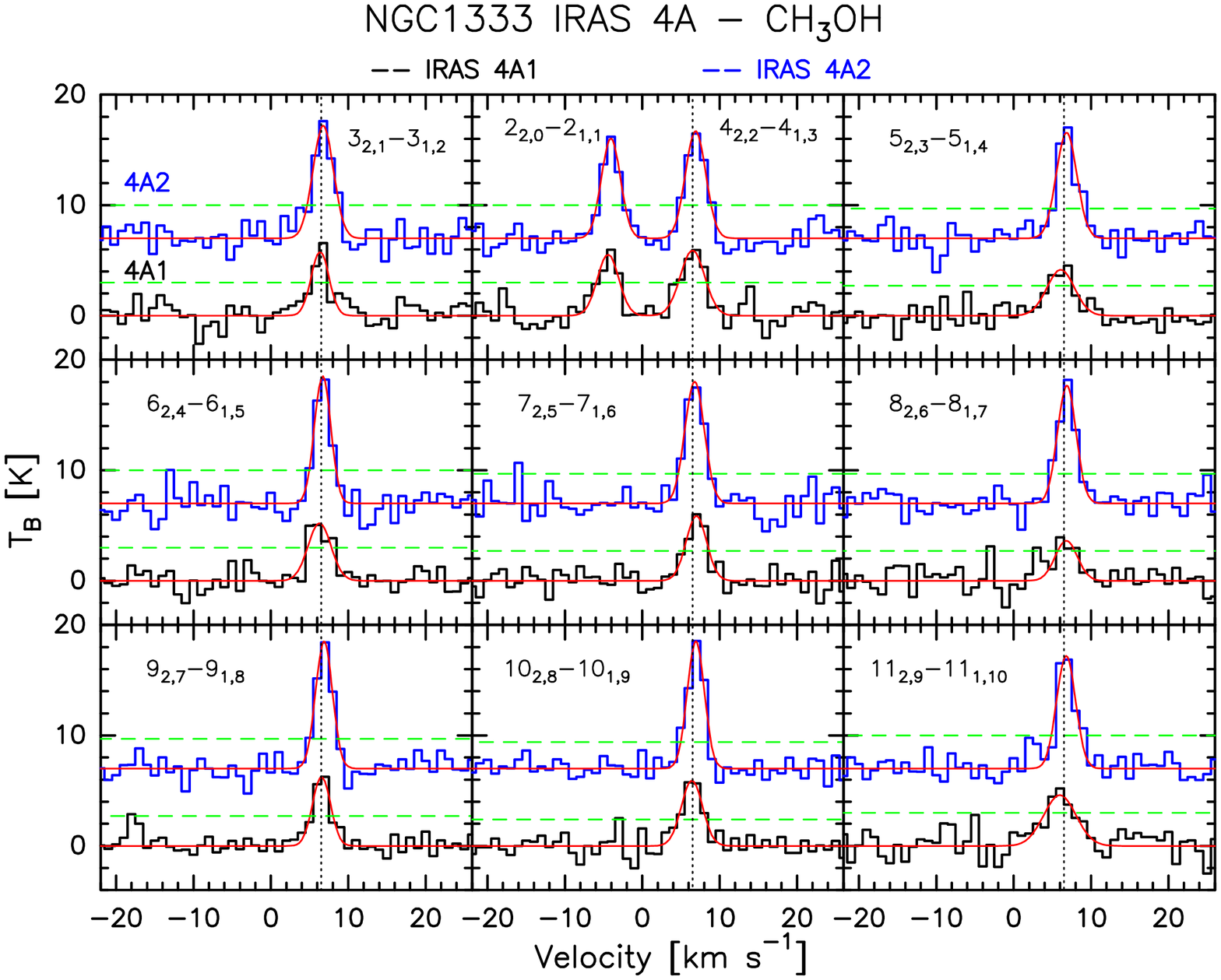}
    \caption{The methanol lines (marked in each panel) detected in the VLA K--band toward 4A1 (black) and 4A2 (blue). The horizontal green dashed lines show the 3$\sigma$ level. The vertical dotted black lines report the $\rm v_{sys}$ (6.7 km s$^{-1}$). The red and magenta curves show the best Gaussian fits (see Table \ref{tab:spectral_params&fit_res}). }
    \label{fig:spectra_ch3oh_iras4a1}
\end{figure*}

All the targeted methanol lines are detected with a signal-to-noise ratio larger than 3 (Table \ref{tab:spectral_params&fit_res}).
Their velocity-integrated spatial distribution is shown in Fig. \ref{fig:methanol_maps+cont}. 
The methanol emission peaks exactly toward the 4A1 and 4A2 continuum peaks, and it is well disentangled, even if unresolved at the current angular resolution, around the two protostars.
 
Figure \ref{fig:spectra_ch3oh_iras4a1} shows the 10 methanol line spectra, isolated and not contaminated by other species, extracted toward the 4A1 and 4A2 continuum peaks.
The lines are slightly brighter toward 4A2 than 4A1, whereas the linewidths are very similar (see also Tab. \ref{tab:spectral_params&fit_res}).
We derived the velocity-integrated line intensities for each detected CH$_3$OH transition using a Gaussian fit, being the profile Gaussian-like. 
The fit results for both sources, namely the integrated emission ($ \int T_b dV$), the linewidth (FWHM), the peak velocities ($V_{peak}$) and the RMS computed for each spectral window, are reported in Table \ref{tab:spectral_params&fit_res}. 
The velocity peaks are consistent with the systemic velocity of the molecular envelope surrounding IRAS 4A
\citep[$\sim6.7$ km s$^{-1}$;][]{choi_high-resolution_2001}.
The line-widths are between 3 and 4 km s$^{-1}$ in agreement with those found by \citet{taquet_constraining_2015} and \citet{lopez-sepulcre_complex_2017} toward 4A2 at mm wavelength. 

In summary, our new VLA observations show a first clear important result: the detection of methanol emission toward 4A1, the protostar where previous mm observations showed no iCOMs emission \citep{lopez-sepulcre_complex_2017}.

\section{Centimeter versus millimeter observations: dust absorption derivation} \label{sec:cm_vs_mm}

We compared our new cm observations of methanol lines with previous ones at 143--146 GHz in order to understand whether the dust absorption, more important at mm than at cm wavelengths, may explain the absence of iCOMs mm line emission in 4A1 \citep{lopez-sepulcre_complex_2017}.
We first carried out a non-LTE analysis of the cm methanol lines from which we derived the gas temperature, density and CH$_3$OH column density toward 4A1 and 4A2 (\S \ref{subsect:T_n_derivation}). Then, using the same parameters, we predicted the methanol line intensities at 143--146 GHz, the frequency of the observations by \citet[][Section \ref{subsect:int_predictions}]{taquet_constraining_2015}. 
Finally, we compared the predicted and measured mm line intensities and we attributed the difference to the absorption of the dust between us and the gas emitting methanol, via the usual equation:
\begin{equation} \label{eq:Absorption}
I^{obs}_\nu=I^{pred}_\nu e^{-\tau_\nu}
\end{equation}
in order to derive the dust optical depth toward 4A1 and 4A2, respectively (Section \ref{sec:dust-abs}).
Please note that the foreground dust opacity obtained by Eq. \ref{eq:Absorption} assumes that the absorbing dust fully covers the emitting gas area, which may not be necessarily the case. Yet, the derived attenuation of the methanol line intensities is still valid, even though it is only an average over the emitting gas area.

\begin{figure}
    \centering
    \includegraphics[scale=0.45]{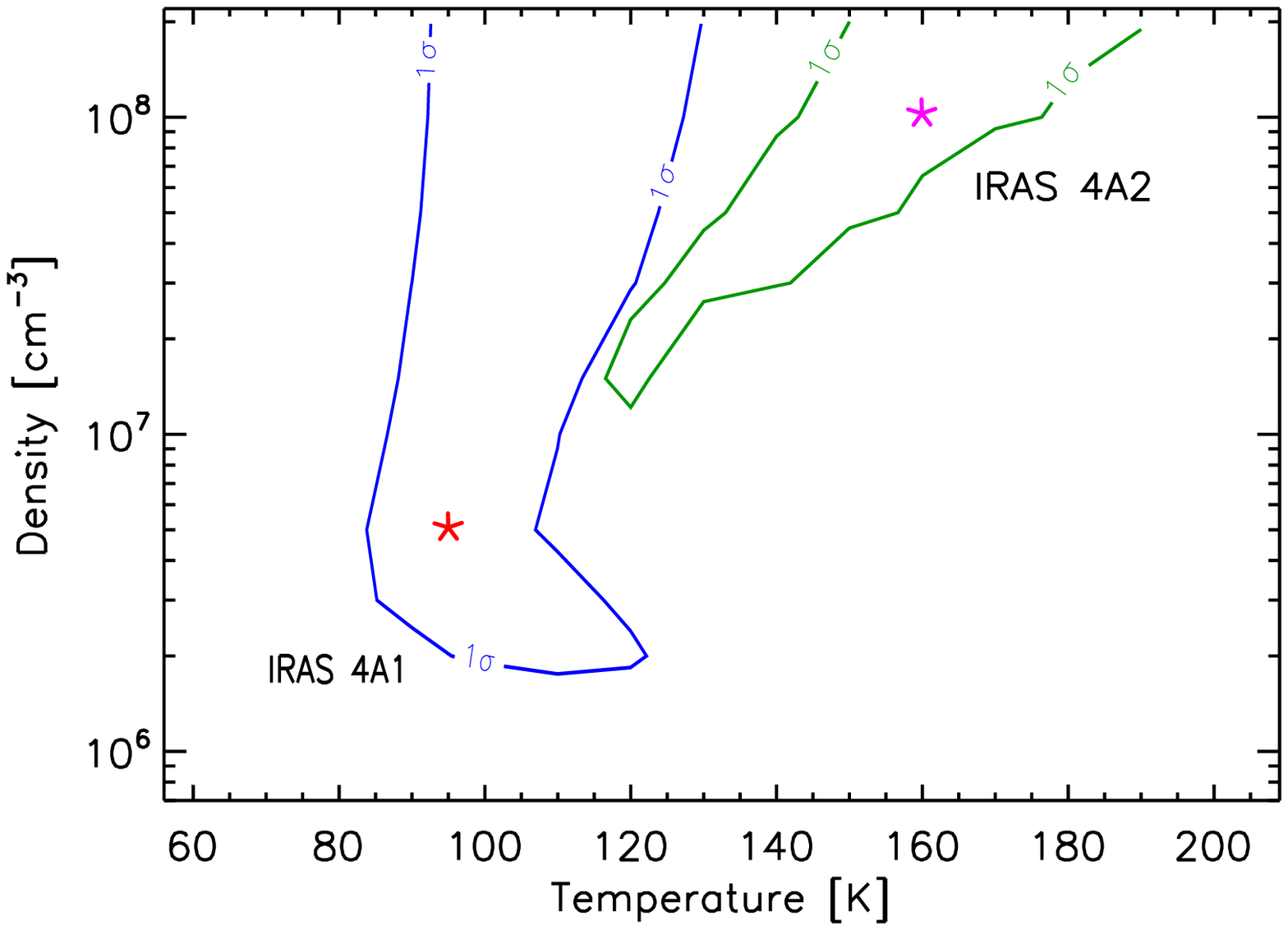}
    \caption{ 
    Density-Temperature $\rm \chi^2$ contour plots.  
    The contours represent 1$\sigma$ confidence level contours for 4A1 (blue) and 4A2 (green), respectively, assuming the best fit values of N$_{\rm CH_3OH}$ and $\theta$ in Table \ref{tab:LVG_results}.
    The best fit solutions are marked by the red (4A1) and magenta (4A2) asterisks. 
    }
    \label{fig:LVG_results}
\end{figure}

\subsection{non-LTE analysis of the cm methanol lines} \label{subsect:T_n_derivation}

To derive the physical properties of the gas emitting CH$_3$OH, namely gas temperature, density and methanol column density, we performed a non-LTE analysis using a Large Velocity Gradient (LVG) code \citet{ceccarelli_theoretical_2003}.
CH$_3$OH can be identified in A- and E-type due to the total spin (I) state of the hydrogen nuclei in the CH$_3$ group: A-type if the total spin function is symmetric (I=3/2), E-type if asymmetric (I=1/2) \citep{rabli_rotational_2010}. 
We used the collisional coefficients of both types of CH$_3$OH with para-H$_2$, computed by \citet{rabli_rotational_2010} between 10 and 200 K for the first 256 levels and provided by the BASECOL database \citep{dubernet_basecol2012:_2013}. 
We assumed a spherical geometry to compute the line escape probability \citep{de_jong_hydrostatic_1980}, the CH$_3$OH-A/CH$_3$OH-E ratio equal to 1, the H$_2$ ortho-to-para ratio equal to 3, and that the levels are populated by collisions and not by the absorption of the dust background photons whose contribution is very likely negligible due to the low values of the CH$_3$OH Einstein coefficients $\rm A_{ij}$.   
Please note that the present LVG analysis only accounts for the line optical depth (to have also the dust $\tau$ in the methanol emitting region would require information on the structure of the region which we do not have, as the emission is unresolved). 

We ran a large grid of models ($\geq$10000) covering the frequency of the observed lines, a total (CH$_3$OH-A plus CH$_3$OH-E) column density $ N_{\rm CH_3OH}$ from $2\times10^{16}$ to $8\times10^{19}$ cm$^{-2}$, 
a gas density $ n_{\rm H_2}$ from $1\times10^{6}$ to $2\times10^{8}$ cm$^{-3}$, and a temperature T from 80 to 200 K.
We then simultaneously fitted the measured CH$_3$OH-A and CH$_3$OH-E line intensities via comparison with those simulated by the LVG model, leaving $ N_{\rm CH_3OH}$, $ n_{\rm H_2}$, $T$, and the emitting size $\theta$ as free parameters.
Following the observations, we assumed the linewidths equal to 3.5 km s$^{-1}$ and 3.0 km s$^{-1}$ for 4A1 and 4A2, respectively, and we included the calibration uncertainty (15\%) in the observed intensities. 

The best fit is obtained for a total CH$_3$OH column density $\rm N_{CH_3OH}=2.8\times10^{19}$ cm$^{-2}$ with reduced chi-square $\rm \chi^2_R=0.6$ for 4A1 and $\rm N_{CH_3OH}=1\times10^{19}$ cm$^{-2}$ with $\rm \chi^2_R=0.1$ for 4A2. 
All the observed lines are predicted to be optically thick and emitted by a source of $0\farcs22$ for 4A1 and $0\farcs24$ for 4A2 ($\sim$70 au) in diameter. 
Solutions with $\rm N_{CH_3OH}\geq1\times10^{18}$ cm$^{-2}$ for 4A2 and $\geq1\times10^{19}$ cm$^{-2}$ for 4A1 are within 1$\sigma$ of confidence level. 
Increasing the methanol column density, the $\chi^2_R$ decreases until it reaches a constant value for $\rm N_{CH_3OH}\geq1\times10^{19}$ cm$^{-2}$ for 4A1 and $\rm N_{CH_3OH}\geq3\times10^{19}$ cm$^{-2}$ for 4A2; this is because all the observed lines become optically thick ($\tau\sim10-30$ for 4A1,$\tau\sim2-6$ for 4A2) and, consequently, the emission becomes that of a black body. The results do not change assuming a linewidth $\pm$0.5 km s$^{-1}$ with respect to the chosen one.

Figure \ref{fig:LVG_results} shows, for both sources, the density-temperature $\chi^2$ surface of the $ N_{\rm CH_3OH}$ best fit. The gas temperature is (90--130) K for 4A1 and (120--190) K for 4A2, while for the gas density we obtained a lower limit of $2\times10^{6}$ cm$^{-3}$ for 4A1 and $1.5\times10^{7}$ cm$^{-3}$ for 4A2, which implies that the levels are LTE populated. 
The fit results are reported in Table \ref{tab:LVG_results}. 
The derived $ n_{\rm H_2}$ and T are consistent with those computed with the model summarised in \citet{su_infall_2019} using our sizes.

\subsection{Predictions of mm methanol line intensities} \label{subsect:int_predictions}

Adopting the 1$\sigma$ range of gas temperature and density derived for 4A1 and 4A2 (Table \ref{tab:LVG_results}), we ran a new grid of LVG models with the CH$_3$OH column density from $1\times10^{18}$ to $8\times10^{19}$ cm$^{-2}$ at 143--146 GHz to predict the methanol line intensities observed by \citet{taquet_constraining_2015}.
We then used the CH$_3$OH $3_1-2_1$ A$^+$ line at 143.866 GHz, which provides the most stringent constraint to the dust optical depth, to compare the predicted intensity with that observed by \citet{taquet_constraining_2015}. 
In the comparison, we took into account our LVG-derived source size and the angular resolution of the \citet{taquet_constraining_2015} observations. 
While for 4A2 we considered the line intensity quoted by \citet{taquet_constraining_2015}, for 4A1, not having CH$_3$OH detection, we used the 3$\sigma$ level of the \citet{taquet_constraining_2015} observations integrated over 3 km s$^{-1}$ (average linewidth toward 4A1: see \S \ref{sec:results}).

The 4A1 and 4A2 CH$_3$OH predicted and observed values are reported in Table \ref{tab:LVG_results}. 
While the two intensities are similar toward 4A2, they differ by about a factor five toward 4A1.

\subsection{Dust absorption toward 4A1 and 4A2}\label{sec:dust-abs}

Assuming that the difference between the predicted and observed intensities is due to the (foreground) dust absorption and using Eq. \ref{eq:Absorption}, we derived the dust optical depth at 143 GHz ($\tau_{\rm dust}^{\rm 143GHz}$; Table \ref{tab:LVG_results}).
While $\rm \tau_{dust}^{\rm 143GHz}$ toward 4A2 is small ($\sim0.3$), that toward 4A1 is large ($\geq1.6$) enough to attenuate the methanol line intensity by a factor $\geq$5.
Therefore, the dust is affecting the mm line emission differently in the two sources.

\begin{table}
    \centering
    \caption{\textit{Top:} Best-fit results and 1$\sigma$ confidence level (range) from the non-LTE LVG analysis of the CH$_3$OH lines toward 4A1 and 4A2. \textit{Bottom:} Comparison of the LVG model predictions with the \citet{taquet_constraining_2015} millimeter observations (see text).}
    \label{tab:LVG_results}
    \hspace{-15mm}
    \resizebox{\columnwidth}{!}{%
    \begin{tabular}{ll|ccccc}
    \hline
    \hline
    \multicolumn{2}{c}{} & \multicolumn{2}{c}{IRAS 4A1} &  \multicolumn{2}{c}{IRAS 4A2}\\
    \hline
    \multicolumn{2}{c|}{} & \multicolumn{4}{c}{LVG Results} \\
    \multicolumn{2}{c|}{} & Best Fit & Range & Best Fit & Range \\
    \hline
    n(H$_2$) & [cm$^{-3}$] & 5$\times10^6$ & $\geq$ 2$\times10^6$ 
    & 1$\times10^8$ & $\geq$1$\times10^7$ \\
    T$_{\rm kin}$ & [K] & 100 & 90-130 &  160 & 120-190 \\
    N$_{\rm CH_3OH}$ & [cm$^{-2}$] & 2.4$\times 10^{19}$ & $\geq$1$\times 10^{19}$ &  1$\times 10^{19}$ & $\geq$1$\times 10^{18}$ \\
    Source size & [$''$] & 0.22 & 0.20-0.24 & 0.24 & 0.22-0.30  \\
    \hline
    \multicolumn{2}{c|}{} & \multicolumn{4}{c}{Predictions vs mm observations} \\
    \hline
    $\rm T_bdV_{pred}$& [K km s$^{-1}$] & \multicolumn{2}{c}{4.7(0.8)} & \multicolumn{2}{c}{9.1(1.2)}\\
    $\rm T_bdV_{obs}$ & [K km s$^{-1}$] & \multicolumn{2}{c}{$\leq$ 0.9} & \multicolumn{2}{c}{6.5(1.9)}\\
    $\rm \tau_{dust}^{143GHz}$ & &  \multicolumn{2}{c}{$\geq$1.6} & \multicolumn{2}{c}{0.3} \\ \hline
    \end{tabular}
    }
\end{table}

\section{Discussion}
\subsection{Is IRAS 4A1 a hot corino?}

So far, only about a dozen hot corinos have been detected (\S \ref{sec:intro}) and the question arises whether this is because they are rare or because the searches have always been carried out at mm wavelengths, where dust could heavily absorb the line emission.

Our first result is that a source that was supposed not to be a hot corino based on mm observations, IRAS 4A1 \citep{lopez-sepulcre_complex_2017}, indeed possesses a region with temperature $\geq100$ K (\S \ref{fig:LVG_results}), namely the icy mantle sublimation one, and shows methanol emission (\S \ref{sec:results}), the simplest of the iCOMs, when observed at cm wavelengths. According to its definition \citep{ceccarelli_hot_2004}, thus, IRAS 4A1 is a hot corino. 

Although we cannot affirm that hot corinos are ubiquitous, it is clear that the searches at mm wavelengths may be heavily biased and that complementary cm observations are necessary to account for dust opacity and understand the occurrence of hot corinos.

\subsection{4A2 versus 4A1: are they chemically different?}

Unlike 4A2, no sign of iCOMs mm emission was revealed toward 4A1 \citep{taquet_constraining_2015,lopez-sepulcre_complex_2017}. 
Using ALMA observations at 250 GHz, \citet{lopez-sepulcre_complex_2017} found that the iCOMs abundances toward 4A2 and 4A1 differ by more than a factor 17, with the largest values ($\sim$100) for HCOOCH$_3$ and CH$_3$CN.

The first question to answer is whether the chemical difference between the two coeval objects is real or due to a different absorption by the surrounding dust.

In Section \ref{sec:dust-abs}, we found that $\rm \tau_{dust}$ at 143 GHz toward 4A1 and 4A2 is $\geq$1.6 and 0.3, respectively (see Table \ref{tab:LVG_results}).
Using the dependence of $\rm \tau_{dust}$ from the frequency ($\rm \tau_{\nu_2}/\tau_{\nu_1}=(\nu_2/\nu_1)^\beta$) and assuming $\beta=2$ (ISM value), the optical depth scaled at 250 GHz \citep[frequency at which][derived the above iCOMs abundance ratios]{lopez-sepulcre_complex_2017} is $\geq$4.9 for 4A1 and 0.9 for 4A2.
Therefore, the different dust absorption toward 4A1 and 4A2 provides us, as lower limit, a factor 55 difference in their line intensities ($\rm I^{A2}/I^{A1}$), comparable to the 4A2/4A1 iCOMs abundance ratios derived by \citet{lopez-sepulcre_complex_2017}. 
A large dust absorption was also suggested by the anomalous flattened continuum spectral index at 100-230 GHz \citep{li_systematic_2017} and the 90$^{\circ}$ flipping of the linear polarization position angles observed at above and below 100 GHz frequencies \citep{ko_resolving_2020}.
 
Although we cannot exclude that a real chemical difference exists between 4A1 and 4A2, the observations so far available cannot support that hypothesis.
Centimeter observations of other iCOMs than methanol are necessary to settle this issue. This conclusion may apply to other binary systems where an apparent chemical difference is observed using mm observations.

\subsection{Are the iCOMs abundances in hot corinos underestimated?}

The dust absorption also affects the iCOMs line intensities in 4A2. 
At 143 GHz $\rm \tau_{dust}$ is 0.3, which leads to underestimate the iCOMs abundances by about 30\%.
At higher frequencies, this factor becomes more important; e.g. at 250 GHz, where several hot corinos studies are carried out (see references in \S \ref{sec:intro}), the absorption factor would be 2.5, and at 350 GHz, frequency where the most sensitive iCOMs search has been carried out \citep[e.g.][]{jorgensen_alma_2016}, the absorption factor would be 6.
This behaviour also agrees with what already found in massive hot cores \citep[e.g.][]{rivilla_formation_2017}.
Therefore, in order to derive reliable iCOMs abundances complementary cm observations are needed to estimate the dust absorption.

\section{Conclusion} \label{sec:conclusion}
We carried out observations of methanol lines at cm wavelengths with the VLA interferometer toward the binary system IRAS 4A, where previous mm observations showed a possible chemical differentiation between the two objects. 
Specifically, while 4A2 showed iCOMs line emission, 4A1 did not.

Our new observations detected ten methanol lines in 4A1 and 4A2 with similar intensities. 
Using a non-LTE analysis and comparing with previous methanol mm observations, we showed that (1) 4A1 is a hot corino, (2) the lack of iCOMs detection toward 4A1 at mm wavelengths is caused by a large dust optical depth, and (3) the determination of the iCOMs abundances toward 4A2 via mm observations is slightly underestimated by the dust absorption.

Therefore, the difficulty in discovering new hot corinos could be because the searches have been carried out at (sub)mm wavelengths, where the dust absorption might be not negligible.
The suspected different chemical nature of coeval objects of the same binary system needs also to be verified at cm wavelengths, as well as the iCOMs abundances estimated from mm observations.

We conclude that centimeter observations of hot corinos are of paramount importance for their correct study.
In the future, next generation instruments in the centimeter wavelenght regime, such as ngVLA \citep{mcguire_science_2018} and SKA \citep{codella_complex_2015}, could be even the most efficient way to identify hot corinos and certainly the most appropriate facilities to study them.

\acknowledgements
    We thank the referee P.T.P Ho for his fruitful comments and suggestions.
    This work has received funding from the European Research Council (ERC) under the European Union's Horizon 2020 research and innovation programme, for the Project ``The Dawn of Organic Chemistry'' (DOC), grant agreement No 741002.
    It was supported by the project PRIN-INAF 2016 The Cradle of Life--GENESIS-SKA (General Conditions in Early Planetary Systems for the rise of life with SKA), and partly supported by the Italian Ministero dell’Istruzione, Università e Ricerca, through the grant Progetti Premiali 2012--iALMA (CUP C52I13000140001). 
    H.B.L. is supported by the Ministry of Science and Technology (MoST) of Taiwan (grant Nos. 108-2112-M-001-002-MY3).
    C.F. acknowledge support from the French National Research Agency in the framework of the Investissements d'Avenir program (ANR-15-IDEX-02), through the funding of the "Origin of Life" project of the Univ. Grenoble-Alpes.

\bibliography{IRAS4A}{}
\bibliographystyle{aasjournal}

\end{document}